\documentclass[11pt]{article}
\usepackage{amsfonts,amsmath}
\usepackage[vcentermath]{youngtab}
\usepackage{young}
\def\bbt{\bibitem}
\def\be{\begin{equation}}
\def\en{\end{equation}}
\def\ber{\begin{eqnarray}}
\def\enr{\end{eqnarray}}
\def\nmb{ \nonumber\\}
\def\d{\partial}

\def\rbrc{\rbrace}
\def\lbrc{\lbrace}
\def\ov{\over }
\def\tld{\tilde}

\def\MTR{Manin triple }

\def\DLG{double Lie group }

\def\Tt{\Theta}
\def\tt{\theta}
\def\sgm{\sigma}

\def\Gm{\Gamma}
\def\im{\imath}

\def\lm{\lambda}

\def\dlt{\delta}
\def\om{\omega}
\def\Om{\Omega}

\begin{document}

\centerline{\bf Generalized K$\ddot{a}$hler Geometry and current algebras}
\centerline{\bf in $SU(2)\times U(1)$ N=2 superconformal WZW model.}
\vskip 1.5 true cm
\centerline{\bf S. E. Parkhomenko}
\centerline{Landau Institute for Theoretical Physics}
\centerline{142432 Chernogolovka,Russia}
\vskip 0.5 true cm
\centerline{spark@itp.ac.ru}
\vskip 0.5 true cm
\centerline{\bf Abstract}
\vskip 0.5 true cm
We examine the Generalized K$\ddot{a}$hler Geometry of quantum N=2 superconformal WZW model on $SU(2)\times U(1)$ and relate the right-moving and left-moving Kac-Moody superalgebra currents to the Generalized K$\ddot{a}$hler Geometry data of the group manifold using Hamiltonian formalism.
\vskip 10pt
 
{\bf 1.} It is well-known that conformal supersymmetric $\sgm$-models with extended supersymmetry are important in the construction
of realistic models of superstring compactification from 10 to 4 dimensions \cite{Gep}. The $\sgm$-model on the 6 -dimensional Calabi-Yau manifold is one of the important examples of the compactification. The Calabi-Yau manifold being a complex K$\ddot{a}$hler manifold is not accidental but caused by a close relation between the extended supersymmetry and K$\ddot{a}$hler geometry. Apart from the metric in a more general
case the background geometry may include an antisymmetric $B$-field. In that case the corresponding 2-dimensional supersymmetric $\sgm$-model have a second supersymmetry when the target-space has a bi-Hermitian geometry, known also as Gates-Hull-Ro$\check{c}$ek geometry \cite{GHR}. In this situation the target manifold contains two complex structures with a Hermitian metric with respect to each of the complex structures. Quite recently it has been shown in \cite{Gualt} that these set of geometric objects, metric, antisymmetric $B$-field and two complex structures antisymmetric with respect to the metric have a unified description in the context of Generalized K$\ddot{a}$hler Geometry (GKG).

 The $N=2$ supersymmetric WZW models on the compact groups \cite{SSTP}, \cite{P}, \cite{QFR} provide a large class of examples of exactly solvable quantum conformal $\sgm$-models whose targets supports simultaneously GKG geometry causing the extended $(2,2)$-supersymmetry and affine Kac-Moody superalgebra structure ensuring the exact solution \cite{KacTod}, \cite{JF}. The GKG nature of the $(2,2)$-supersymmetry in these models has been studied in the series of works \cite{SevT}, \cite{RSSev}, \cite{HLRUZ}, 
\cite{SevSW}, \cite{SevST}. The relation between Generalized Complex geometry and Kac-Moody superalgebra symmetry has already been discussed in papers \cite{HelZ}, \cite{HelZ1}. In this paper I undertake more thorough study of this relation using WZW model on the group $SU(2)\times U(1)$ as an example. In a more general context it would be also important to see if there are GKG targets which allow the $W$-superalgabras conserved currents. Perhaps Kazama-Suzuki coset models \cite{KS1}-\cite{KS3} can be related to such targets according to the results presented in \cite{P1}.

 The approach I follow is close to the one used in papers \cite{Z}, \cite{BLPZ}, \cite{Zlec} where the Hamiltonian formalism of $N=1$ supersymmetric $\sgm$-model has been applied to show a relation between the extended supersymmetry in phase-space and generalized complex structure of the target space. I intend to show that Hamiltonian approach works as well
to relate GK Geometry and Kac-Moody superalgebra structure for the case of $N=2$ superconformal WZW model on a compact group manifold. It is because the right-moving and left-moving Kac-Moody superalgebra currents are given by the right and left Poisson-Lie group actions coming from GK Geometry of the group manifold. It also allows us to give Poisson-Lie interpretation of Wakimoto formulas.


\vskip 10pt
{\bf 2.} It is convenient to start examination with the supersymmetric WZW model on an arbitrary compact group $G$ using superfield formalism. The super world-sheet of the model is parametrized by the light-cone
even coordinates
$\sgm_{\pm}$, and odd coordinates $\tt_{\pm}$. The super-derivatives are given by
\ber
D_{\pm}={\d\ov\d\tt^{\pm}}+\tt^{\pm}\d_{\pm}, \ \d_{\pm}={\d\ov \d \sgm_{0}}\pm{\d\ov \d\sgm_{1}}
\label{2.lconeD}
\enr
The superfield $g(\sgm_{\pm},\tt^{\pm})$ takes values in the compact group $G$. The action of the model \cite{swzw} is given by
\ber
S= -k(\int d^{2}\sgm d^{2} \tt(<g^{-1}D_{+}g,g^{-1}D_{-}g>   
\nmb
         -\int d^{2}\sgm d^{2}\tt dt
          <g^{-1}\frac{\d g}{\d t},\lbrc g^{-1}D_{-}g,g^{-1}D_{+}g\rbrc>)
\label{2.action}
\enr 
The classical equations of motion are nothing else but the conservation low equations:
\ber
 D_{-}(g^{-1}D_{+}g)=D_{+}(D_{-}gg^{-1})= 0.
\label{2.meq}
\enr

{\bf 3.} The outhor of papers \cite{P},\cite{QFR} established a correspondence between extended
supersymmetric WZNW models and finite-dimensional Manin triples. By the definition ~\cite{Drinf}, a \MTR $({\bf g},{\bf g_{+}},{\bf g_{-}})$
consists of a Lie algebra ${\bf g}$, with nondegenerate invariant inner product
$<,>$ and isotropic Lie subalgebras ${\bf g_{\pm}}$ such that
${\bf g}={\bf g_{+}}\oplus {\bf g_{-}}$ as a vector space. 
There is a one to one correspondence \cite{P} between a complex Manin triple endowed with antilinear involution which conjugates isotropic subalgebras $\tau: {\bf g_{\pm}}\rightarrow {\bf g_{\mp}}$ and
a complex structure on a real Lie algebra of the compact group that ensures connection between Manin triple construction of  $N=2$   Virasoro superalgebra currents of \cite{P},\cite{QFR} and approach \cite{SSTP} based on the complex structure on the Lie algebra. Thus, for an arbitrary compact Lie group $G$ with the complex structure $J$
the complexification ${\bf g^{\mathbb{C}}}$ of the Lie algebra ${\bf g}$ of the group has the Manin triple
structure $({\bf g^{\mathbb{C}}},{\bf g_{+}},{\bf g_{-}})$ where the isotropic subalgebras are the $\pm\im$-eigenspaces
of $J$. The Lie group version
of this triple is the \DLG $(G^{\mathbb{C}},G_{+},G_{-})$
~\cite{SemTian}, \cite{LuW}. 
The real Lie group $G$ is extracted
from its complexification by hermitian conjugation $\tau$:
\ber
G= \lbrc g\in G^{\mathbb{C}}|\tau (g)=g^{-1}\rbrc       
\label{3.realG}
\enr
Because of $G^{\mathbb{C}}=G_{+}G_{-}$ each element $g\in G^{\mathbb{C}}$
admits two decompositions
\ber
g= g_{+}g^{-1}_{-}= {\tld g}_{-}{\tld g}^{-1}_{+},  
\label{3.decomp}
\enr
where ${\tld g}_{\pm}$ are dressing transformed
elements of $g_{\pm}$ ~\cite{LuW}.
Taking into account (\ref{3.realG}) and (\ref{3.decomp}) we conclude that the
element $g$ from $G^{\mathbb{C}}$ belongs to $G$ iff
\ber
\tau (g_{\pm})= {\tld g}^{-1}_{\mp}      
\label{3.unitar}
\enr
These equations mean that there is an left and right dressing actions of the complex groups $G_{\pm}$ on $G$ \cite{SemTian} so the elements of $G$ can be parametrized
by the elements from the complex group $G_{+}$ (or $G_{-}$).

 More generaly one has to consider the
set $W$ of classes $G_{+}\backslash G^{\mathbb{C}}/ G_{-}$
and pick up a representative $w$ for each class $[w]\in W$.
It gives the stratifications (Bruhat decomposition)
of $G^{\mathbb{C}}$ ~\cite{AlMal}:
\ber
G^{\mathbb{C}}= \bigcup_{[w]\in W} G_{+}wG_{-}=
\bigcup_{[w]\in W} G_{-}wG_{+}
\label{3.Bruhat}
\enr
Thus, in a general case the compact group $G$ can be represented as a set of dressing action orbits endowed with natural complex coordinates \cite{P2}. In this note we will not look into this complication.

{\bf 4.}The simplest nonabelian compact group with complex structure is $SU(2)\times U(1)$. Its complexification is $GL(2,\mathbb{C})$. Let us fix the following basis in the Lie algebra $gl(2,\mathbb{C})$
\ber
e^{0}={1\ov 2}\left(\begin{array}{cc}
                 1+\im &0\\
                 0&-1+\im
                 \end{array}\right), \
e^{1}=\left(\begin{array}{cc}
                 0 &1\\
                 0& 0
                 \end{array}\right),
\nmb
e_{0}={1\ov 2}\left(\begin{array}{cc}
                 1-\im &0\\
                 0&-1-\im
                 \end{array}\right),
\
e_{1}=\left(\begin{array}{cc}
                 0&0\\
                 1&0
                 \end{array}\right)
\label{4.basis}
\enr
The isotropic subalgebras forming the corresponding Manin triple $(gl(2,\mathbb{C}),{\bf g}_{+}, {\bf g}_{-})$ are given by
\ber
{\bf g_{+}}=\mathbb{C} e_{0}\oplus \mathbb{C} e_{1}, \ {\bf g_{-}}=\mathbb{C} e^{0}\oplus \mathbb{C} e^{1}
\label{4.Mtriple}
\enr
where the invariant scalar product $<,>$ is given by trace so that $<e^{a},e_{b}>\equiv Tr(e^{a}e_{b})=\dlt^{a}_{b}$.
The first decomposition from (\ref{3.decomp}) takes the following form
\ber
g_{+}=\left(\begin{array}{cc}
                 \exp ({1\ov 2}(1-\im)x^{0})     &0\\
                 \exp ({1\ov 2}(1-\im)x^{0})x^{1}&\exp (-{1\ov 2}(\im+1)x^{0})
                 \end{array}\right),
								\nmb
g_{-}=\left(\begin{array}{cc}
                 \exp ({1\ov 2}(\im+1)y^{0})&\exp ({1\ov 2}(\im-1)y^{0})y^{1}\\
                 0                       &\exp ({1\ov 2}(\im-1)y^{0})
                 \end{array}\right)
\label{4.U2decomp}
\enr
The solution of (\ref{3.unitar}) for the group $SU(2)\times U(1)$ is given by
\ber
y^{0}=\bar{x}^{0}+\ln (1+|x^{1}|^{2}), \ y^{1}=\exp(-x^{0}+\bar{x}^{0})\bar{x}^{1} 
\nmb								
g_{+}g^{-1}_{-}=(1+|x^{1}|^{2})^{-{1+\im\ov 2}}a^{1-\im}\bar{a}^{-1-\im}
\left(\begin{array}{cc}
                 1&-a^{-2}\bar{a}^{2}\bar{x}^{1}\\
                 x^{1}&a^{-2}\bar{a}^{2}
                 \end{array}\right)\in SU(2)\times U(1)
\label{4.U2decomp1}
\enr
where $a=\exp ({1\ov 2}x^{0})$.
Hence, the $x^{0}, x^{1}$ are the holomorphic coordinates on $SU(2)\times U(1)$ determined by the 
left dressing action of $G_{+}$. One can show \cite{P2} that these coordinates are holomorphic with respect to the complex structure $J_{r}$ generated by the right translations from the complex structure of the Manin triple
$(gl(2,\mathbb{C}),{\bf g}_{+}, {\bf g}_{-})$. 

 We employ these coordinates to rewrite the $WZW$action (\ref{2.action}) on $SU(2)\times U(1)$ in the form of the supersymmetric $\sgm$-model action
\ber
S={1\ov 2}\int d^{2}\sgm d^{2} \tt E_{ij}\rho^{i}_{+}\rho^{j}_{-}
\label{4.action}
\enr
Here, the matrix $E_{ij}$ depends on the superfields $X^{i}$, $\bar{X}^{i}$ corresponding to the coordinates 
$x^{i}$, $\bar{x}^{i}$. It is related \cite{P2} to the Semenov-Tian-Shansky simplectic form $\Om_{ij}$:
\ber
E_{ij}=\im \Om_{ik}(J_{r})^{k}_{j},
\nmb
\Om=-{k\ov 2}(1+|x^{1}|^{2})^{-1}(\bar{x}^{1}\rho^{0}\wedge \rho^{1}+\rho^{0}\wedge\rho^{\bar{0}}+
\rho^{1}\wedge\rho^{\bar{1}}-x^{1}\rho^{\bar{0}}\wedge \rho^{\bar{1}})
\label{4.STSform}
\enr
The superfields $\rho^{i}_{\pm}=(\rho^{a}_{\pm},\rho^{\bar{a}}_{\pm})$ are the world-sheet restrictions of the holomorphic
$\rho^{a}=Tr(e^{a}dg_{+}g_{+}^{-1})$ and anti-holomorphic $\rho^{\bar{a}}=Tr(e^{a}d\bar{g}_{+}\bar{g}_{+}^{-1})$ 1-forms w.r.t $J_{r}$. They are given by
\ber
\rho_{\pm}^{0}=D_{\pm}X^{0}, \ \rho_{\pm}^{1}=D_{\pm}X^{1}+X^{1}D_{\pm}X^{0},
\nmb
\rho_{\pm}^{\bar{0}}=D_{\pm}\bar{X}^{0}, \ \rho_{\pm}^{\bar{1}}=D_{\pm}\bar{X}^{1}+\bar{X}^{1}D_{\pm}\bar{X}^{0}
\label{4.hol1forms}
\enr
The target space metric $g_{ij}$ and $B$-field $B_{ij}$ can be read off from $E_{ij}$:
\ber
g_{ij}={1\ov 2}(E_{ij}+E_{ji}), \ B_{ij}=-{1\ov 2}(E_{ij}-E_{ji})
\label{4.gB}
\enr
The relations (\ref{4.STSform}), (\ref{4.gB}) are nothing else but the Theorem from \cite{HLRUZ} describing GKG data in terms of gerbes.

{\bf 5.}  Now we find the canonically conjugated momentum and super-Hamiltonian following the analisis of \cite{BLPZ}.
Let us define new set of odd world-sheet coordinates and derivatives
\ber
\tt^{+}={1\ov\sqrt{2}}(\tt^{0}+\tt^{1}), \ \tt^{-}={\im\ov\sqrt{2}}(\tt^{1}-\tt^{0}),
\nmb
D_{0}={\d\ov\d\tt^{0}}+\tt^{0}\d_{1}+\tt^{1}\d_{0},
\
D_{1}={\d\ov\d\tt^{1}}+\tt^{1}\d_{1}+\tt^{0}\d_{0}
\label{5.newsuperderiv}
\enr
so that $D_{0}^{2}=D_{1}^{2}=\d_{1}$, $D_{0}D_{1}+D_{1}D_{0}=2\d_{0}$.

 Integrating out $\tt^{0}$-variable the action (\ref{4.action}) becomes
\ber
S=\int d^{2}\sgm d\tt^{1}(\d_{0}X^{\mu}X^{*}_{\mu}-{1\ov 2}g_{\mu\nu}\d_{1}X^{\mu}D_{1}X^{\nu}
\nmb
-{1\ov 2}(X^{*}_{\mu}-B_{\mu\lm}D_{1}X^{\lm})g^{\mu\nu}D_{1}(X^{*}_{\nu}-B_{\nu\eta}D_{1}X^{\eta})-
\nmb
{1\ov 2}D_{1}X^{\mu}\Gm^{\tau}_{\mu\nu}g^{\nu\lm}(X^{*}_{\lm}-B_{\lm\eta}D_{1}X^{\eta})(X^{*}_{\tau}-
B_{\tau\om}D_{1}X^{\om})
\nmb
+{1\ov 2}H_{\mu\nu\lm}D_{1}X^{\mu}D_{1}X^{\nu}g^{\lm\eta}(X^{*}_{\eta}-B_{\eta\tau}D_{1}X^{\tau})
\nmb
-{1\ov 6}H^{\mu\nu\lm}(X^{*}_{\mu}-B_{\mu\eta}D_{1}X^{\eta})(X^{*}_{\nu}-B_{\nu\tau}D_{1}X^{\tau}))
(X^{*}_{\lm}-B_{\lm\om}D_{1}X^{\om}))
\label{5.HamiltonB}
\enr
where 
\ber
X^{*}_{\mu}=g_{\mu\nu}D_{0}X^{\nu}+B_{\mu\nu}D_{1}X^{\nu}
\label{5.momentumB}
\enr
is canonically conjugated momentum to the field $X^{\mu}$, 
$\Gm^{\lm}_{\mu\nu}$ are the Christofell symbols of the Levi-Civita connection for the metric $g_{\mu\nu}$:
\ber
g_{\mu\eta}\Gm^{\eta}_{\nu\lm}={1\ov 2}(g_{\mu\nu,\lm}+g_{\lm\mu,\nu}-g_{\nu\lm,\mu})
\label{D.Christof}
\enr
and
\ber
H_{\mu\nu\lm}={1\ov 2}(B_{\mu\nu,\lm}+B_{\lm\mu,\nu}+B_{\nu\lm,\mu})
\label{D.WZW3form}
\enr
determines WZW 3-form.
From (\ref{5.HamiltonB}) we can also read off the density of the super-Hamiltonian which defines an evolution
along the variables $(\sgm_{0},\tt^{0})$

 Quantization of the canonical Poisson brackets for the superfields $X^{\mu}$, $X^{*}_{\nu}$ can be represented as the singular part of the following OPE
\ber
X^{*}_{\mu}(Z_{1})X^{\nu}(Z_{2})=Z_{12}^{-{1\ov 2}}\dlt^{\nu}_{\mu}\hbar+reg.
\label{5.canonicope}
\enr
where the new supercoordinate $Z=(z,\Tt)$, $z=\exp(\im\sgm_{1})$, $\Tt=(\im z)^{1\ov 2}\theta_{1}$ on the super-circle is introduced and $Z_{12}=z_{1}-z_{2}-\Tt_{1}\Tt_{2}$.

 In what follows we rescale the canonical variables by
$X^{*}_{\mu}(Z)\rightarrow {1\ov \hbar}X^{*}_{\mu}(Z)$, $X^{\mu}(Z)\rightarrow X^{\mu}(Z)$
to get the standard OPE
\ber
X^{*}_{\mu}(Z_{1})X^{\nu}(Z_{2})=Z_{12}^{-{1\ov 2}}\dlt^{\nu}_{\mu}+reg.
\label{5.canonicope1}
\enr

{\bf 6.} The main feature of the superconformal WZW model is that it has two copies of Kac-Moody superalgebra symmeties generated by the left-moving $D_{+}gg^{-1}$ and right-moving $g^{-1}D_{-}g$ currents. These two algebras determine the dynamics 
and $N=2$ supersymmetry of the model completely due to the generalized Sugawara construction \cite{SSTP}, \cite{P}, \cite{QFR} (see also \cite{Getz}). Therefore it is important to get these currents in the Hamiltonian approach and then recover the Hamiltonian density as well as the total $N=(2,2)$ Virasoro superalgebra. 

 The classical conserved left-moving currents are given by
\ber
L^{a}\equiv Tr(e^{a}(D_{0}+D_{1})gg^{-1}), \ L_{a}\equiv Tr(e_{a}(D_{0}+D_{1})gg^{-1}), a=0,1
\label{6.L}
\enr
In what follows we will use the common notation $L_{i}$ for $(L^{a}$, $L_{a})$. Using (\ref{4.U2decomp1}) one finds that
\ber
L_{i}=\Om_{ij}(g^{jk}(\rho^{*}_{k}-B_{kn}\rho^{n})+\rho^{j})=\Om_{ij}g^{jk}(\rho^{*}_{k}+E_{kn}\rho^{n})
\label{6.Lcurrent}
\enr
here $\rho^{i}=(Tr(e^{a}D_{1}g_{+}g_{+}),Tr(e_{a}D_{1}g_{+}g_{+}))$ are the super-circle restrictions of the corresponding 1-forms and $\rho^{*}_{i}$ are the conjugated to $\rho^{i}$ superfields:
\ber
\rho^{*}_{0}=X^{*}_{0}-X^{1}X^{*}_{1}, \ \rho^{*}_{1}=X^{*}_{1}
\nmb
\rho^{*}_{\bar{0}}=\bar{X}^{*}_{0}-\bar{X}^{1}\bar{X}^{*}_{1}, \ \rho^{*}_{\bar{1}}=\bar{X}^{*}_{1}
\label{6.rhodual}
\enr
Introducing the Semenov-Tian-Shansky Poisson bivector \cite{SemTian}
\ber
P\equiv \Om^{-1}={2\ov k}(X^{1}\rho^{*}_{0}\rho^{*}_{1}+\rho^{*}_{0}\bar{\rho}^{*}_{0}+\rho^{*}_{1}\bar{\rho}^{*}_{1}-
\bar{X}^{1}\rho^{*}_{\bar{0}}\rho^{*}_{\bar{1}})
\label{6.PSTS}
\enr
and lifting up the indeces in (\ref{6.Lcurrent}) one gets
\ber
L^{i}=-{k\ov 2}(P(\rho^{i})+\im J_{r}\rho^{i})
\label{6.LPoisson}
\enr
Notice that formula close to (\ref{6.LPoisson}) has been obtained in the context of topological $\sgm$-models \cite{AlStr}.
The expression (\ref{6.LPoisson}) gives Poisson-Lie interpretation for the left-moving current algebra in $N=2$ superconformal WZW model. Rewriting (\ref{6.LPoisson}) by the fields $X^{i}(Z)$, $X^{*}_{i}(Z)$ we find that it is nothing else but the direct sum of Wakimoto realizations \cite{Wak}
\ber
L^{1}=X^{*}_{1}-\bar{X}^{1}\bar{X}^{*}_{0}+(\bar{X}^{1})^{2}\bar{X}^{*}_{1}-{k\ov 2}D\bar{X}^{1}-
{k\ov 2}\bar{X}^{1}D\bar{X}^{0},
\nmb
L^{0}=-X^{1}X^{*}_{1}+X^{*}_{0}+\bar{X}^{1}\bar{X}^{*}_{1}-{k\ov 2}D\bar{X}^{0}
\nmb
L_{0}=-X^{1}X^{*}_{1}-\bar{X}^{*}_{0}+\bar{X}^{1}\bar{X}^{*}_{1}+{k\ov 2}DX^{0}
\nmb
L_{1}=-(X^{1})^{2}X^{*}_{1}+X^{1}X^{*}_{0}-\bar{X}^{*}_{1}+{k\ov 2}DX^{1}+{k\ov 2}X^{1}DX^{0}
\label{6.Wakimot}
\enr
The quantum versions of (\ref{6.LPoisson}), (\ref{6.Wakimot}) differ from the classical ones by the normal ordering and reproduce the correct OPE's due to (\ref{5.canonicope1}):
\ber
L^{0}(Z_{1})L_{0}(Z_{2})=Z_{12}^{-1}k+reg., 
\nmb
L^{1}(Z_{1})L_{1}(Z_{2})=Z_{12}^{-1}k+Z_{12}^{-{1\ov 2}}(L^{0}+L_{0})(Z_{2})+reg.,
\nmb
L^{0}(Z_{1})L_{1}(Z_{2})=-Z_{12}^{-{1\ov 2}}L_{1}(Z_{2})+reg., \
L_{0}(Z_{1})L^{1}(Z_{2})=Z_{12}^{-{1\ov 2}}L^{1}(Z_{2})+reg.
\nmb
L^{0}(Z_{1})L^{1}(Z_{2})=Z_{12}^{-{1\ov 2}}L^{1}(Z_{2})+reg., \
L_{0}(Z_{1})L_{1}(Z_{2})=-Z_{12}^{-{1\ov 2}}L_{1}(Z_{2})+reg.,
\label{6.LcurrentOPE}
\enr
Recovering the Plank constant for the currents (\ref{6.Wakimot}) one sees that the level $k$ is scaled
as ${1\ov \hbar}$.

 The classical conserved right-moving currents are given by
\ber
R^{a}\equiv Tr(e^{a}g^{-1}(D_{0}+D_{1})g), \ R_{a}\equiv Tr(e_{a}g^{-1}(D_{0}+D_{1})g), a=0,1
\label{6.R}
\enr
Introducig the common notation $R_{i}$ for the currents $(R^{a}$, $R_{a})$ one can obtain
\ber
R^{i}=-{k\ov 2}(P(\tld{\rho}^{i})-\im J_{l}\tld{\rho}^{i})
\label{6.RPoisson}
\enr
where $\tld{\rho}^{i}=(Tr(e^{a}D_{1}\tld{g}_{+}\tld{g}_{+}),Tr(e_{a}D_{1}\tld{g}_{+}\tld{g}_{+}))$
are the super-circule restrictions of holomorphic and anti-holomorphic 1-forms w.r.t. the complex structutre $J_{l}$ which is generated by the left translations on the group $SU(2)\times U(1)$. The group element $\tld{g}_{+}$ is determined by the second decomposition from (\ref{3.decomp}) and parametrized by the dressing transformed coordinates 
$\tld{x}^{0}$, $\tld{x}^{1}$ which are related to the $x^{0}$, $x^{1}$ by
\ber
\tld{x}^{0}=-x^{0}-\ln(1+|x^{1}|^{2}), \ \tld{x}^{1}=-\exp (x^{0}-\bar{x}^{0})x^{1}
\label{6.tldx}
\enr
We also have 
\ber
\tld{\rho}^{0}=D_{1}\tld{X}^{0}, \ \tld{\rho}^{1}=D_{1}\tld{X}^{1}+\tld{X}^{1}D_{1}\tld{X}^{0},
\
\tld{\rho}^{\bar{a}}=\bar{\tld{\rho}}^{a}
\label{6.tldhol1forms}
\enr
where $\tld{X}^{i}$ are the superfields corresponding to the coordinates $\tld{x}^{i}$.
 The expression (\ref{6.RPoisson}) implies that the canonicaly conjugated superfields $\tld{X}^{*}_{i}$ (as well as the fields $\tld{\rho}^{*}_{i}$) have been introduced
so that their commutation relations coincide with (\ref{5.canonicope1}) after the quantization.
Then, one can see that quantum versions of the currents (\ref{6.RPoisson}) written in terms of these new canonical variables
$\tld{X}^{i}(Z)$, $\tld{X}^{*}_{i}(Z)$ nearely match with the Wakimoto formulas (\ref{6.Wakimot}). The only difference is that the level of the right-moving algebra is equal $-k$ and this is the only case when the right-moving currents commute to the left ones \cite{Regrep}. 

 To summarize, the expressions (\ref{6.LPoisson}),  (\ref{6.RPoisson}) determine the left-moving and right-moving current superalgebras in terms of GKG data of the group manifold. It is natural to expect that they are valid
for an arbitrary $N=2$ supersymmetric WZW model. 
 
{\bf 7.} Now we reproduce the Sugawara construction for the left-moving and right-moving $N=2$ Virasoro superalgebra currents from GKG data: $\sgm$-model metric $g_{ij}$, B-field $B_{ij}$ and complex structures $J_{l,r}$. 

 Let us define two generalized complex structures
\cite{BLPZ}, \cite{SevST} acting in the direct sum of tangent and cotangent bundles on the group:
\ber
{\bf J}_{1}={1\ov 2}\left(\begin{array}{cc}
1  & 0 \\
B & 1 \\
\end{array}\right)\left(\begin{array}{cc}
J_{l}+J_{r}     & -\om^{-1}_{l}+\om^{-1}_{r} \\
\om_{l}-\om_{r} & -J^{T}_{l}-J^{T}_{r}       \\
\end{array}\right)
\left(\begin{array}{cc}
1  & 0 \\
-B  & 1 \\
\end{array}\right)
\nmb
{\bf J}_{2}={1\ov 2}\left(\begin{array}{cc}
1  & 0 \\
B & 1 \\
\end{array}\right)\left(\begin{array}{cc}
J_{l}-J_{r}     & -\om^{-1}_{l}-\om^{-1}_{r} \\
\om_{l}+\om_{r} & -J^{T}_{l}+J^{T}_{r}       \\
\end{array}\right)
\left(\begin{array}{cc}
1  & 0 \\
-B  & 1 \\
\end{array}\right)
\label{7.GCstruct}
\enr
where $\om_{l,r}=gJ_{l,r}$, $\om^{-1}_{l,r}=-J_{l,r}g^{-1}$. 
According to \cite{BLPZ}, \cite{SevST} the left-moving $K_{l}(Z)$ and right-moving $K_{r}(Z)$ $U(1)$ supercurrents of the $N=2$ Virasoro superalgebra are given by
\ber
K_{l}=\im<(DX,X^{*}),({\bf J}_{1}+{\bf J}_{2})(DX,X^{*})>=
\nmb
\im(\om_{l}^{-1})^{ij}(X^{*}_{i}+(g-B)_{ik}DX^{k})
(X^{*}_{j}+(g-B)_{jn}DX^{n})
\nmb
K_{r}=\im<(DX,X^{*}),({\bf J}_{1}-{\bf J}_{2})(DX,X^{*})>=
\nmb
-\im(\om_{r}^{-1})^{ij}(X^{*}_{i}-(g+B)_{ik}DX^{k})(X^{*}_{j}-(g+B)_{jn}DX^{n})
\label{7.U1currents}
\enr
Using (\ref{4.STSform}), (\ref{6.PSTS}) as well as the fact that $\rho^{a}$, $\tld{\rho}^{a}$, $a=0,1$ are holomorphic
1-forms w.r.t. $J_{r}$, $J_{l}$ respectively one can deduce
\ber
K_{l}={\im\ov 2k}\om_{l}(L,L),
\
K_{r}=-{\im\ov 2k}\om_{r}(R,R)
\label{7.classicKlr}
\enr
where we introduced Lie algebra valued currents $L=L_{a}e^{a}+L^{a}e_{a}$ and $R=R_{a}e^{a}+R^{a}e_{a}$.

 Quantization of these expressions may lead to quantum corrections, but the result is known due to \cite{P}, \cite{QFR}, (see also \cite{Getz}):
\ber
K_{l}(Z)={1\ov k}(L^{a}L_{a}+f_{a}DL^{a}-f^{a}DL_{a}),
\
K_{r}(Z)=-{1\ov k}(R^{a}R_{a}+f_{a}DR^{a}-f^{a}DR_{a})
\label{7.LRquantumK}
\enr
where the normal ordering is implied, $f_{a}=f_{ab}^{b}$, $f^{a}=f^{ab}_{b}$ and $f_{ab}^{c}$, $f^{ab}_{c}$ are the structure constants of the isotropic subalgebras ${\bf g_{\pm}}$. The geometric meaning of the quantum corrections was found in \cite{HelZ}, \cite{HelZ1}.

 One can check the commutation relations for the modes of the currents (\ref{7.LRquantumK}) and find remaining generators of $N=2$ Virasoro superalgebra using the realization \cite{KacTod}.  We will begin looking into this realization for the currents $L^{a}(Z)$, $L_{a}(Z)$:
\ber
L^{a}=\sqrt{k}\psi^{a}+\Tt (j^{a}+f^{ab}_{c}\psi^{c}\psi_{b}-{1\ov 2}f_{bc}^{a}\psi^{b}\psi^{c}),
\nmb
L_{a}=\sqrt{k}\psi_{a}+\Tt (j_{a}+f_{ab}^{c}\psi_{c}\psi^{b}-{1\ov 2}f^{bc}_{a}\psi_{b}\psi_{c})
\label{7.KacTod}
\enr
where $\psi^{a}(z)$, $\psi_{a}(z)$ are the free fermionic currents and $j^{a}(z)$, $j_{a}(z)$ are the bosonic currents determined by the OPE: 
\ber
\psi^{a}(z_{1})\psi_{b}(z_{2})=z_{12}^{-1}\dlt^{a}_{b}+...
\nmb
j^{a}(z)j^{b}(w)=-(z-w)^{-2}{1\ov 2}(e^{a},e^{b})
          +(z-w)^{-1}f^{ab}_{c}j^{c}(w)+reg,   
					\nmb
          j_{a}(z)j_{c}(w)=-(z-w)^{-2}{1\ov 2}(e_{a},e_{b})
          +(z-w)^{-1}f_{ab}^{c}j_{c}(w)+reg,   
					\nmb
          j^{a}(z)j_{b}(w)=(z-w)^{-2}{1\ov 2}
          (2k\delta^{a}_{b}-(e^{a},e_{b}))      
          +(z-w)^{-1}(f_{bc}^{a}j^{c}-f^{ac}_{b}j_{c})(w)+reg, 
\label{7.KacTodope}
\enr
where $(,)$ is a Killing form. Taking into account (\ref{7.KacTodope}) the currents (\ref{7.KacTod}) satisfy (\ref{6.LcurrentOPE})
thus it is easy to deduce
\ber
K_{l}=I(z)+\Tt{1\ov\sqrt{2}}(G^{+}-G^{-})(z),
\nmb
I(z)=\psi^{a}\psi_{a}+{1\ov k}f_{a}(j^{a}+f^{ab}_{c}\psi^{c}\psi_{b})-{1\ov k}f^{a}(j_{a}+f_{ab}^{c}\psi_{c}\psi^{c}),
\nmb
G^{+}=\sqrt{2\ov k}(j^{a}\psi_{a}-{1\ov 2}f^{ad}_{c}\psi_{a}\psi_{d}\psi^{c}),
\nmb
G^{-}=\sqrt{2\ov k}(j_{a}\psi^{a}-{1\ov 2}f_{ad}^{c}\psi^{a}\psi^{d}\psi_{c})
\label{7.QuantumKGG}
\enr
We see that even part $I(z)$ of $K_{l}$ neither more nor less than the $U(1)$-current of $N=2$ Virasoro superalgebra while the odd contribution is given by spin-${3\ov 2}$ currents $G^{\pm}$ of the superalgebra. It corresponds with the expressions found in \cite{P}, \cite{QFR}, \cite{Getz}. 

 The right-moving currents of $N=2$ Virasoro superalgebra can be obtained analogously. Thus we recovered all currents of the $N=(2,2)$ Virasoro superalgbra and hence, the density of the quantum super-Hamiltonian from the GK Geometry of $SU(2)\times U(1)$ group manifold. We also showed that left-moving and right-moving Kac-Moody superalgebra symmetries in quantum $N=2$ superconformal WZW model are determined by left and right Poisson-Lie group actions on the group manifold and lead to the Wakimoto-type formulas. Taking into account that our analysis has local nature it would be important to consider global construction of the Kac-Moody superalgebra currents in terms of GKG data. Perhaps the most appropriate way to do that is to use gerbe descrition of GK Geometry found in \cite{HLRUZ}.

\vskip 10pt
\centerline{\bf ACKNOWLEDGEMENTS}
\frenchspacing
The work was performed with the financial support of the Russian Science Foundation
(Grant No.14-12-01383).

\vfill
\end{document}